\documentclass[11pt]{article}
\usepackage{mathrsfs}
\usepackage{amsmath}
\usepackage{amsfonts}
\usepackage{amstext}
\usepackage{amscd}
\usepackage{amsthm}
\usepackage{setspace}

\newcommand{\Asla}{\ensuremath{\ A\hspace{-0.6em}/}}

\newcommand{\Dsla}{\ensuremath{\ D\hspace{-0.7em}/}}

\newcommand{\Psl}{\ensuremath{\ P\hspace{-0.7em}/}}

\newcommand{\R}{\ensuremath{\mathbb{R}}}

\numberwithin{equation}{section}
\begin{document}

\title{\textbf{Paramagnetism, zero modes and mass singularities 
in QED in $1+1$, $2+1$ and $3+1$ dimensions}}
\author{M. P. Fry\thanks{Electronic address: mpfry@maths.tcd.ie} \\
\textit{School of Mathematics, Trinity College, Dublin 2,
Ireland. }}
\date{July 1996}
\maketitle

\begin{abstract}
The interplay of paramagnetism, zero modes of the Dirac operator and
fermionic mass singularities on the fermionic determinants in quantum
electrodynamics in two, three and four dimensions is discussed.
\\ \\ \\ \\ \\ \\ \\ 
PACS number(s): 12.20.Ds, 11.10.Kk, 11.15.Tk
\vspace{5ex}
\end{abstract}

\pagebreak

\setcounter{section}{1}
\begin{center}
\large{\bf I.\@ INTRODUCTION}
\end{center}

One of the oldest outstanding problems of nonperturbative QED is the
calculation of its fermionic determinant.  This determinant is obtained by integrating
over the fermionic degrees of freedom in the presence of a background
potential $A_\mu$, thereby producing the one-loop, gauge invariant effective 
action $\ln\det(F_{\mu\nu})$ that appears in the calculation of all
physical processes. The problem is that since the determinant is itself
part of a functional integral over $A_\mu$, it has to be known for all
fields.  If the gauge field is given an infrared cutoff then $A_\mu$ is
concentrated on $\mathscr{S}',$ the Schwartz space of real tempered 
distributions.  Consequently, $\det$ is needed for fields that can be rough
and that have no particular symmetry.  Nevertheless, modulo certain technical
assumptions, bounds can be placed on the Euclidean determinants in 
QED$_2$, QED$_3$ and QED$_4$ for particular classes
of fields.  These bounds are reviewed in [1].  Here we wish to discuss
what additional insight can be gained by paying particular attention to the
interplay of the paramagnetic tendency of charged fermions in external
magnetic and electric fields, zero modes of the Dirac operator 
$\hspace{-0.3em}\Dsla = \hspace{-0.5em}\Psl - \hspace{-0.5em} \Asla$,
and fermionic mass singularities.  Recall that in
Euclidean space $\mathbf{E}$ and $\mathbf{B}$ are on the same footing, so that
one may speak of paramagnetism in the presence of an electric field as well.

The effect of paramagnetism is particularly transparent in Schwinger's
proper time definition of the fermionic determinant [2--4],
\begin{equation}
\ln \mbox{det}_{ren} = \frac{1}{2}\int_0^{\infty}
\frac{dt}{t}
\left\{ \mbox{Tr}
\left( e^{ - P^2t} 
 -\exp [ - ( D^2 + \tfrac{1}{2} \sigma^{\mu\nu}F_{\mu\nu} )
t \right) + \frac{\|F\|^2}{24\pi^2} \right \} e^{-tm^2}.
\end{equation}
Here $D^2=(P-A)^2$; $m$ is the unrenormalized fermion mass; $\sigma^{\mu\nu}=
(1/2i)[\gamma^\mu,\gamma^\nu],\gamma^{\mu\dagger} = -\gamma^\mu$, and
$\|F\|^2=\int d^4xF^2_{\mu\nu}(x)$.  The procedure for smoothing the rough
fields and for introducing a volume cutoff is described in refs.\ [1,5].  
Here we will just assume that $A_\mu$ and $F_{\mu\nu}$ are smooth and that
 $F_{\mu\nu}$ is square-integrable.  The last item in (1.1) is the second-order
mass shell charge renormalization subtraction required for the small $t$ limit
of the integral to converge.  The same definition may be used in lower
dimensions by omitting the charge renormalization subtraction and dropping
the subscript ``ren'' on the determinant.

Paramagnetism is immediately manifest through the square-integrable zero modes
and zero energy threshold resonance states of the nonnegative Pauli operator
$D^2+\frac{1}{2}\sigma F$ for suitable $A_\mu$.  By inspection of definition 
(1.1), these tend to drive $\ln \mbox{det}_{ren}$ toward negative values
and will introduce nonperturbative mass singularities.

In two and three dimensions paramagnetism extends on up the spectrum
of the Pauli operator by lowering its eigenvalues on average relative
to those of the free Hamiltonian $P^2$, as manifested by the bound
$\ln\det \leq 0$ [3,6,7]. There is no such bound in four dimensions [3]
due to the positive sign of the charge renormalization counterterm and,
hence, ultimately due to the nonasymptotic freedom of QED$_4$.  It is of
interest to quantify this interplay between  nonasymptotic freedom and
paramagnetism as this bears on the stability of  QED$_4$.  
For if $\ln \mbox{det}_{ren}$ grows faster than a
quadratic in the field strength then it is doubtful that it is integrable
with respect to the potential's gauge-fixed Gaussian measure. These
introductory remarks will be developed in Sec.\ III.\@  In Sec.\ II we take up
the case of QED$_2$ and conclude with a discussion of QED$_3$ in Sec.\ IV.

\setcounter{section}{2}
\setcounter{equation}{0}
\begin{center}
\large{\bf II.\@ ZERO MODES IN QED$_2$} 
\end{center}

For Euclidean QED in two-dimensional space-time, otherwise known as the
massive Schwinger model, Eq.\ (1.1) gives
\begin{eqnarray}
\frac{\partial}{\partial m^2}\ln\mbox{det}_{QED2}=\frac{1}{2}
\mbox{Tr} \left[(D^2-\sigma_3 B +m^2)^{-1} -(P^2+m^2)^{-1} \right],
\end{eqnarray}
where $B=F_{01}=\partial_0A_1-\partial_1A_0$, and the coupling $e$ has been
absorbed into $A_\mu$ so that the flux $\Phi=\int d^2r B({\mathbf r})$
is dimensionless.  Thus, $\ln\mbox{det}_{QED2}$ is determined by the
quantum mechanics of a charged fermion confined to a plane in the presence
of a static magnetic field perpendicular to the plane.  As noted in Sec.\ I,
this field and its associated potential are assumed sufficiently smooth with
enough fall off so that everything done here makes mathematical sense.  It
was conjectured in [8] (see Eq.\ (5.67) and below) that
\begin{eqnarray}
\lim_{m^2=0} m^2 \frac{\partial}{\partial m^2} \ln\mbox{det}_{QED2}=
\frac{|\Phi|}{4\pi}.
\end{eqnarray}
Referring to (2.1), this would indicate that the leading mass singularity
of $\ln\mbox{det}_{QED2}$ is determined by the zero modes of the Pauli
operator $(P-A)^2-\sigma_3B$.  The number of square-integrable zero modes is
$[ | \Phi | / 2 \pi]$, all with $\sigma_3=1 \; (\sigma_3=-1)$ if $\Phi > 0  \;
(\Phi < 0)$,
where [x] stands for the nearest integer less than $x$ and [0] = 0 [9].  The
remaining fractional part of $\Phi$ is related to the zero energy scattering
phase shifts for $-\hspace{-1 ex}\Dsla^{\hspace{0.2 ex}2}$ [10].  Here
we will 
prove (2.2) for the 
case
$B({\mathbf r}) \geq 0$ or $B({\mathbf r}) \leq 0$.  But first note that (2.2)
contradicts the perturbative result
\begin{eqnarray}
\lim_{m^2=0} m^2 \frac{\partial}{\partial m^2} \ln\mbox{det}_{QED2}=
\frac{\Phi^2}{8 \pi^2}+O(\Phi^4),
\end{eqnarray}
obtained by expanding (1.1) in the field strength. In fact, the series in (2.3)
is not even asymptotic.  The reason is clear:
square-integrable zero modes and zero-energy threshold resonances cannot
be captured by a perturbative expansion. Eqs.\ (2.2) and (2.3) stand as a
salutary lesson on how badly perturbation theory can go wrong in QED.

Now for the proof.  Suppose $B({\mathbf r}) \geq 0$.  From Eqs.\ (5) and (6)
of [11] one has
\begin{eqnarray}
m^2 \frac{\partial}{\partial m^2} \ln\mbox{det}_{QED2} & = & 
\frac{\Phi}{4 \pi} +m^2 \mbox{Tr}[(D^2+B+M^2)^{-1}-(P^2+m^2)^{-1}] \nonumber
\\
& \leq & \frac{\Phi}{4 \pi} -\frac{m^2}{4\pi}\int d^2 r 
\ln \left(1+\frac{B({\mathbf r})}{m^2} \right),
\end{eqnarray}
where the trace in the first line of (2.4) is over space indices only.  Since
\begin{eqnarray}
m^2 \int d^2 r \ln \left(1+\frac{B({\mathbf r})}{m^2} \right) \leq \int d^2 B({\mathbf r})
= \Phi,
\end{eqnarray}
for $m^2 \geq 0$, the integral in (2.4) converges uniformly for $m^2 \geq 0$.
Therefore the limit $m^2=0$ may be taken inside the integral, giving
\begin{eqnarray}
\lim_{m^2=0} m^2 \frac{\partial}{\partial m^2} \ln\mbox{det}_{QED2} \leq
\frac{\Phi}{4 \pi}.
\end{eqnarray}
Next, note that for $M^2 \geq m^2$,
\begin{eqnarray}
\mbox{Tr} & \hspace{-5.5 ex} [ & \hspace{-5.5ex} (D^2 + B +m^2)^{-1} - (P^2+m^2)^{-1}] 
\nonumber \\
& \geq & \mbox{Tr}[(D^2 + B +M^2)^{-1} - (P^2+m^2)^{-1}] \nonumber \\
& = & \mbox{Tr}[(D^2 + B +M^2)^{-1} - (P^2+M^2)^{-1}] \nonumber \\
& & \hspace{5.5 ex} + \mbox{Tr}[(P^2+M^2)^{-1}-(P^2+m^2)^{-1}] \nonumber \\
& = & \mbox{Tr}[(D^2 + B +M^2)^{-1} - (P^2+M^2)^{-1}] + \pi \ln (\tfrac{m^2}{M^2}).
\end{eqnarray}
Thus, (2.7) and the first line of (2.4) imply
\begin{eqnarray}
\lim_{m^2=0} m^2 \frac{\partial}{\partial m^2} \ln\mbox{det}_{QED2}
\geq \frac{\Phi}{4 \pi},
\end{eqnarray}
and hence (2.6) and (2.8) together imply (2.2).
The case $B({\mathbf r}) \leq 0$ is dealt with along the same lines, the only
change being that $B \rightarrow -B \geq 0$ and $\Phi \rightarrow -\Phi \geq 0$
in (2.4) following ref.\ [11], thereby establishing (2.2) for both cases.

From (2.2) one gets the small mass limit
\begin{eqnarray}
\ln\mbox{det}_{QED2}=\frac{|\Phi |}{4 \pi} \ln m^2 + R(m^2),
\end{eqnarray}
where $\underset{m^2=0}{\lim} (R(m^2)/ \ln m^2)=0$.  Letting $A_\mu(x) \rightarrow \lambda
A_\mu(\lambda x)$, $F_{\mu\nu}(x) \rightarrow \lambda^2 F_{\mu\nu}(\lambda x),
\, \lambda > 0$, in (1.1) there results the exact scaling relation
\begin{eqnarray}
\ln\mbox{det}(\lambda^2 F_{\mu\nu}(\lambda x),m^2)=\ln\mbox{det}(F_{\mu\nu}(x),
m^2/ \lambda^2),
\end{eqnarray}
which holds in two, three and four dimensions.  Applying (2.10) to (2.9) gives
\begin{eqnarray}
\ln\mbox{det}_{QED2} (\lambda^2 B (\lambda {\mathbf r}), m^2) 
\underset{\lambda >>1}{\sim}
- \frac{ | \Phi |}{2 \pi} \ln \lambda +\mathscr{R}(\lambda),
\end{eqnarray}
where $\underset{\lambda=\infty}{\lim} (\mathscr{R}(\lambda) / \ln\lambda) = 0$.  Note the
minus sign, indicating paramagnetism. Also note that the coefficient of $\ln\lambda$ is
the absolute value of the anomaly. However, this still does not answer the main 
question: how does $\ln\mbox{det}_{QED2}$ grow for large field strength obtained
by the scaling $B({\mathbf r}) \rightarrow \lambda B({\mathbf r}), \:
\lambda > 0$? 
In [8], Eq.\ (5.67), we conjected
\begin{eqnarray}
\lim_{\lambda=\infty}(\ln\mbox{det}_{QED2} \: (\lambda B,m^2)/ \lambda\ln\lambda)=
-\frac{ | \Phi | }{2 \pi}.
\end{eqnarray}

\setcounter{section}{3}
\setcounter{equation}{0}
\begin{center}
\large{\bf III.\@ PARAMAGNETISM, ZERO MODES AND MASS SINGULARITIES IN QED$_4$} 
\end{center}

From (1.1),
\begin{eqnarray}
m^2 \frac{\partial}{\partial m^2} \ln\mbox{det}_{ren}=
\frac{m^2}{2} \mbox{Tr} \left[(D^2+\tfrac{1}{2}\sigma
F+m^2)^{-1}-(P^2+m^2)^{-1} \right] 
-\frac{ \| F \|^2}{48 \pi^2}.
\end{eqnarray}
The main question here is: What is the $m^2=0$ limit of (3.1)?  The answer is
not so straightforward as in the case of QED$_2$.  Since $D\hspace{-1.6 ex}/ $ and $\gamma_5$
anti-commute, then in a basis in which $\gamma_5$ is diagonal, 
$\gamma_5=\bigl( \begin{smallmatrix}
1&0\\0&-1
\end{smallmatrix} \bigr)$,  $D\hspace{-1.5 ex}/ $ takes the supersymmetric
form [12]
\begin{eqnarray}
D\hspace{-1.5 ex}/=\left(\begin{matrix}0&\Delta_-\\ \Delta_+&0 
\end{matrix} \right),
\end{eqnarray}
with $\Delta^\dagger_-=-\Delta_+$, and hence
\begin{eqnarray}
-D^2\hspace{-2.5 ex}/ \:=D^2+\tfrac{1}{2}\sigma F 
=\left(\begin{matrix} H_+ &0\\0&H_- \end{matrix} 
\right),
\end{eqnarray}
where
\begin{eqnarray}
H_\pm=(P-A)^2-\boldsymbol{\sigma}\cdot({\mathbf B} \pm {\mathbf E}).
\end{eqnarray}
The global anomaly for  $D\hspace{-1.6 ex}/ $ may be put in the form [10]
\begin{eqnarray}
\frac{1}{4 \pi^2} \int d^4x \: {\mathbf E}\cdot{\mathbf B}(x) & = &
\underset{m^2=0}{\lim} \mbox{Tr}
\left(\frac{m^2}{H_++m^2}-\frac{m^2}{H_-+m^2}\right) \nonumber \\
& = & n_+ -n_- + \frac{1}{\pi} \sum_l \mu(l) \left(\delta^l_+(0)-\delta^l_-(0) \right),
\end{eqnarray}
where $n_\pm$ are the number of square integrable zero modes of $H_\pm$;
$\delta^l_\pm(0)$ are the scattering phase shifts for $H_\pm$ as the
energy tends to zero; $l$ is a degeneracy parameter such as angular
momentum, and $\mu(l)$ is a weight factor. The last line in (3.5) is a
generalization of the Atiyah-Singer index theorem [13] to non-compact 
manifolds.

We will now use (3.1) and (3.5) to rederive the asymptotic form of the
constant field result for $\ln\mbox{det}_{ren}$.  Although it is a
rederivation, a new physical insight is gained that goes well beyond the
constant field case.

By making two rotations (corresponding to a Lorentz boost and a rotation
in Minkowski space) a frame can be found in which ${\mathbf E}$ and
${\mathbf B}$ are parallel.  Select the frame where ${\mathbf B}'=(0,0,B')$,
 ${\mathbf E}'=(0,0,E')$, with
\begin{eqnarray}
B' &=& \frac{1}{2} (| {\mathbf B} +{\mathbf E}| + | {\mathbf B} -{\mathbf E}|) 
\\
E' &=& \frac{1}{2} (| {\mathbf B} +{\mathbf E}| - | {\mathbf B} -{\mathbf E}|),
\end{eqnarray}
and of course ${\mathbf B'}\cdot{\mathbf E'}={\mathbf B}\cdot{\mathbf E}$.  Suppose 
${\mathbf B}\cdot{\mathbf E}>0$.  Then $B'$, $E'>0$ and $H_\pm$ take the form
\begin{eqnarray}
H_\pm=(P-A)^2-\sigma_3(B'\pm E').
\end{eqnarray}
Choose $A_x=-yB'$, $A_z=-tE'$, $A_t=A_y=0$ so that
\begin{eqnarray}
(P-A)^2=H_{E'} +H_{B'},
\end{eqnarray}
where $H_{E'}$, $H_{B'}$, describe two oscillators with energies $(2n+1)E'$,
$(2m+1)B'$ with $m,n=0,1$,...  By inspection of (3.8), all zero modes have
positive chirality.  Likewise, when ${\mathbf B}\cdot{\mathbf E}<0$ then
$B'>0$ and $E'<0$, in which case we conclude that all zero modes have
negative chirality.  Because there are no scattering states, the phase shifts
are zero in (3.5).

Referring back to (3.1), suppose ${\mathbf E}\cdot{\mathbf B}>0$.  Then the
fact that there are no negative chirality zero modes combined with (3.5)
gives
\begin{eqnarray}
\underset{m^2=0}{\lim} m^2 \frac{\partial}{\partial m^2}\ln\mbox{det}_{ren}
& = & \frac{1}{2} \, \underset{m^2=0}{\lim} m^2 \mbox{Tr} 
[(H_++m^2)^{-1}-(H_-+m^2)^{-1}] \nonumber \\
& + & \! \! \underset{m^2=0}{\lim}m^2 \mbox{Tr}[(H_-+m^2)^{-1}-(P^2+m^2)^{-1}] -
\frac{ \|F\|^2}{48 \pi^2} \nonumber \\
& = & \frac{{\mathbf E}\cdot{\mathbf B}V}{8 \pi^2}-
\frac{(E^2+B^2)V}{24 \pi^2},
\end{eqnarray}
where $V$ is the volume of the space-time box, and the trace operation
includes a trace over a $2 \times 2$ spin space.  Similarly, when
${\mathbf E}\cdot{\mathbf B} <0$ there are no positive chirality zero modes
and so (3.5) gives
\begin{eqnarray}
\underset{m^2=0}{\lim} m^2 \frac{\partial}{\partial m^2}\ln\mbox{det}_{ren}
& = &  -\frac{{\mathbf E}\cdot{\mathbf B}V}{8 \pi^2} -
\frac{(E^2+B^2)V}{24 \pi^2}.
\end{eqnarray}
Hence we can conclude that for small $m^2$,
\begin{eqnarray}
\ln\mbox{det}_{ren}=-\frac{V}{24 \pi^2}(B^2+E^2-3|{\mathbf E}\cdot{\mathbf B}|)
\ln m^2 + R(m^2),
\end{eqnarray}
using the same definition of $R$ as in Sec.\ II.\@ From the scaling relation
(2.10) we obtain for $\lambda >>1$,
\begin{eqnarray}
\ln\mbox{det}_{ren}(\lambda^2 F_{\mu\nu}(\lambda x),m^2)
=\frac{V}{24 \pi^2}(B^2+E^2-3|{\mathbf E}\cdot{\mathbf B}|)
\ln \lambda^2 + \mathscr{R}(\lambda),
\end{eqnarray}
where $\mathscr{R}(\lambda)$ is defined in Sec.\ II.\@  This is the same result
that is obtained by scaling the exact constant field expression for
$\ln\mbox{det}_{ren}$ [2,4].  

On the basis of (3.13) it is tempting to make the general field conjecture
\begin{eqnarray}
\underset{\lambda=\infty}{\lim}(\ln\mbox{det}_{ren}(\lambda^2 F_{\mu\nu}
(\lambda x),m^2)/ \ln \lambda^2)=\frac{1}{8} \beta_1 \|F\|^2 - \frac{1}{2}|A|,
\end{eqnarray}
where $\beta_1$ is obtained from the Callan-Symanzik function $\beta=2\alpha
/ 3\pi + O(\alpha^2)$, i.e.\, $\beta_1=1/6\pi^2$, and $A$ is the anomaly,
$\frac{1}{16\pi^2} \int d^4x F^\star_{\mu\nu} F_{\mu\nu}$, with
$ F^\star_{\mu\nu}=\frac{1}{2} \epsilon_{\mu\nu\alpha\beta}F_{\alpha\beta}$.
Recall that $\|F\|^2 \geq 16 \pi^2 |A|$ so that the right-hand side of
(3.14) is greater than or equal to $-|A|/6$.

Whether or not (3.14) is correct, this much is now clear.  Paramagnetism
results in zero modes of $D^2 \hspace{-2.5 ex}/$, driving 
$\ln\mbox{det}_{ren}$ toward negative values that tend to offset the
nonasymptotic freedom of QED$_4$ that enters through the positive sign
of the charge renormalization subtraction.  The zero modes contribute to
the leading mass singularity of $\ln\mbox{det}_{ren}$ which can be related to
its strong field behavior by scaling.  As a consequence of paramagnetism
and associated zero modes the sign of $\ln\mbox{det}_{ren}$ for large
field strength is not definite, and so the question of QED$_4$'s stability
becomes all the more interesting.

In order to make further progress one has to go beyond constant fields.
For example, self-dual (anti-self-dual) fields ${\mathbf B}=-{\mathbf E}
\:({\mathbf B}={\mathbf E})$ will, by inspection of (3.4), have only
negative (positive) chirality square-integrable zero modes.  It would
be helpful if one knew the rules, if any, that allow one to predict
such results in more general cases.

\setcounter{section}{4}
\setcounter{equation}{0}
\begin{center}
\large{\bf IV.\@ ZERO MODES IN QED$_3$ }
\end{center}

Making the continuation of QED$_{2+1}$ to Euclidean space, the planar
electric field combines with the normal magnetic field to form a three
dimensional static magnetic field ${ \mathbf B}({\mathbf r})$ on $\R^3$.
Here we will use the $2 \times 2$ Dirac matrices $\gamma^\mu=
(i\sigma_1,i\sigma_2,i\sigma_3)$. Then (1.1) gives 
\begin{eqnarray}
m^2 \frac{\partial}{\partial m^2} \ln\mbox{det}_{QED3} = \frac{m^2}{2} \mbox{Tr}
[(( {\mathbf P}-{\mathbf A})^2 - \boldsymbol{\sigma}\cdot{\mathbf B}+m^2)^{-1}
-(P^2 +m^2)^{-1}].
\end{eqnarray}
Again we ask the question: What is the $m^2=0$ limit of (4.1)?  Of course the
answer is zero for a constant magnetic field [14], or indeed any unidirectional
magnetic field, since the momentum parallel to the field serves as an
extra infrared cutoff.  
Remarkably, it was not known until 1986 whether the equation
\begin{eqnarray}
[({\mathbf P}-{\mathbf A})^2 - \boldsymbol{\sigma}\cdot{\mathbf B}] \psi = 0,
\end{eqnarray}
had any bound state solutions for ${\mathbf B} \in L^2(\R^3)$, and hence
for ${\mathbf A} \in L^6(\R^3)$ when  ${\mathbf A}$ is in the Coulomb
gauge $\boldsymbol{\nabla}\cdot{\mathbf A}=0$ [15,16].  Matters 
are not helped by the absence of an index theorem in three dimensions.  It
is now known that bound states exist and that their degeneracy can be large,
with an upper bound growing like $\int d^3r | {\mathbf B}({\mathbf r})|^{3/2}$
[16].  This will be derived below.

Since it is now known that bound state solutions to (4.2) exist, (4.1) tells
us that $\ln\mbox{det}_{QED3}$ has a logarithmic mass singularity whose
coefficent can grow as fast as $\int | {\mathbf B} |^{3/2}$.  By scaling,
Eq.\ (2.10), we expect
\begin{eqnarray}
\ln\mbox{det}_{QED3} (\lambda^2 {\mathbf B} (\lambda {\mathbf r}), m^2)
\underset{\lambda >>1}{\sim}
-c \int d^3 r |  {\mathbf B} ({\mathbf r}) |^{3/2} \ln \lambda^2,
\end{eqnarray}
where $c$ is a positive constant, again reflecting paramagnetism.  Therefore,
the strong field ``folk theorem'' that 
$\ln\mbox{det}_{QED3} \sim - \int |{\mathbf B}|^{3/2}$ is probably
wrong.  
One wonders if the coefficient of the 
logarithm in (4.3) has a topological 
origin as in two dimensions.

Since the upper bound, $c \int  |  {\mathbf B}|^{3/2}$, on the number of
bound state solutions of (4.2) is not derived in [16] it might be helpful
to provide a derivation here.  It is suggested in [16] that a variational principle be
used together with the inequality $| (\psi, \boldsymbol{\sigma}\cdot{\mathbf B}
\psi) | \leq (\psi, |{\mathbf B}| \psi)$.  Accordingly, the ground state
energy is given by
\begin{eqnarray}
E \: (=0) \, & = & \inf_{ \| \psi \| = 1} (\psi, [({\mathbf P}-{\mathbf A})^2-
\boldsymbol{\sigma}\cdot{\mathbf B}]\:\psi) \nonumber \\
& \geq & \inf_{ \| \psi \| = 1} (\psi, [({\mathbf P}-{\mathbf A})^2-
|{\mathbf B}|]\:\psi) \nonumber \\
& \geq & \inf_{ \| \psi \| = 1} (|\psi|, [({\mathbf P}^2-|{\mathbf B}|]\:|\psi|) 
\nonumber \\
& \geq & \inf_{ \| \phi \| = 1} (\phi, [({\mathbf P}^2-|{\mathbf B}|]\:\phi).
\end{eqnarray}
The third line in (4.4) follows from Simon's variant of Kato's inequality
expressing the diamagnetism of spinless bosons [17].  By (4.4), the number
of bound states of the Hamiltonian $P^2-|{\mathbf B}|$ provides an upper
bound on the number of bound states of the Pauli Hamiltonian 
$({\mathbf P}-{\mathbf A})^2-\boldsymbol{\sigma}\cdot{\mathbf B}$.  The
number $N$ of bound states for the Hamiltonian $P^2 - |{\mathbf B}|$
can be estimated by the Cwikel-Lieb-Rozenblum bound [18]
\begin{eqnarray}
N \leq 0.1156 \int d^3 r | {\mathbf B}({\mathbf r}) |^{3/2},
\end{eqnarray}
which is the result of [16] with the addition of a precise constant.

\begin{center}
\bf{References}
\end{center}
\begin{description}
\item [{[1]}] M. P. Fry, ``QED in inhomogeneous magnetic fields'', U. Dublin,
Trinity College preprint, hep-th/9606037.
\item [{[2]}] J. Schwinger, Phys.\ Rev.\ {\bf 82}, 664(1951).
\item [{[3]}] E. Seiler, in {\it Gauge Theories:  Fundamental
Interactions and Rigorous Results}, Proceedings of the International
School of Theoretical Physics, Poiana Brasov, Romania, 1981, edited by
P. Dita, V. Georgescu, and P. Purice, Progress in Physics Vol.\ 5
(Birkh\"{a}user, Boston, 1982), p.263.
\item [{[4]}] W. Dittrich and M. Reuter, {\it Effective Lagrangians in
Quantum Electrodynamics},  Lecture Notes in Physics Vol.\ 220 (Springer,
Berlin, 1985).
\item [{[5]}] M. P. Fry, Phys.\ Rev.\ D{\bf 45}, 682(1992).
\item [{[6]}]  D. Bridges, J. Fr\"{o}hlich, and E. Seiler, Ann.\ Phys.\
(N.Y.) {\bf 121}, 227(1979).
\item [{[7]}] D. H. Weingarten, Ann.\ Phys.\ (N.Y.) {\bf 126},
154(1980).
\item [{[8]}] M. P. Fry, Phys.\ Rev.\ D{\bf 51}, 810(1995).
\item [{[9]}] Y. Aharonov and A. Casher, Phys.\ Rev.\ A{\bf 19}, 2461(1979).
\item [{[10]}] R. Musto, L. O'Raifeartaigh, and A. Wipf,  Phys.\ Lett.\
B{\bf 175}, 433(1986).
\item [{[11]}] M. P. Fry, Phys.\ Rev.\ D{\bf 53}, 980(1996).
\item [{[12]}] E.  Witten, Nucl.\ Phys.\ B{\bf 188}, 513(1981).
\item [{[13]}] M. Atiyah and I. Singer, Ann.\ Math.\ {\bf 87}, 485(1968); {\bf 87},
 546(1968).
\item [{[14]}] A.N. Redlich, Phys.\ Rev.\ D{\bf 29}, 2366(1984).
\item [{[15]}] J. Fr\"{o}hlich, E. Lieb, and M. Loss, Commun.\ Math.\ Phys.\ 
{\bf 104}, 251(1986).
\item [{[16]}] M. Loss and H. T. Yau, Commun.\ Math.\ Phys.\ 
{\bf 104}, 283(1986).
\item [{[17]}] B. Simon, Phys.\ Rev.\ Lett.\ {\bf 36}, 1083(1976).
\item [{[18]}] E. Lieb, Proc.\ Amer.\ Math.\ Soc.\ Symposia in
Pure Math.\ {\bf 36}, 241(1980);
 E. H. Lieb and W. E. Thirring, in {\it Studies in
Mathematical Physics, Essays in Honor of Valentine Bargmann}, edited by E. H.
Lieb, B. Simon, and A. S. Wightman (Princeton U. Press, Princeton,
1976), p.\ 269.
\end{description}

\end{document}